\documentclass[12pt]{article}
\usepackage[utf8]{inputenc}
\usepackage{amsmath}
\usepackage{xcolor}
\usepackage{booktabs}
\usepackage{fullpage}
\usepackage{todonotes}
\newcommand{\JCM}[1]{{\protect \todo[inline, color = orange]{JCM --- #1}}}

\usepackage{tikz}
\tikzset{
BasicNode/.style={circle, draw= black!50, fill=colora!40, thin, minimum size
  = 5mm, inner sep = 0mm}}

\usetikzlibrary{arrows.meta}

\usepackage{xpatch}

\definecolor{colora}{RGB}{0,115,179}
\definecolor{colorb}{RGB}{230,154,0} 
\definecolor{colorc}{RGB}{0,154,128} 
\definecolor{colord}{RGB}{205,10,179}
\definecolor{colore}{RGB}{255,32,0}
\definecolor{colorf}{RGB}{240,228,66}
\definecolor{colorg}{RGB}{90,179,230}
\definecolor{colorh}{RGB}{205,154,179}

\definecolor{colorS}{RGB}{0,154,128}
\definecolor{colorI}{RGB}{255,32,0}
\definecolor{colorR}{RGB}{205,10,179}
\definecolor{colorE}{RGB}{240,228,66} 

\usepackage[noend]{algpseudocode}
\usepackage[nothing]{algorithm}
\algrenewcommand{\algorithmicrequire}{\textbf{Input:}}
\algrenewcommand{\algorithmicensure}{\textbf{Output:}}
\algnewcommand{\algAND}{\textbf{and}}
\algrenewcommand{\textproc}{\textsf}

\newtheorem{lemma}{Lemma}
\newcommand{\erdosrenyi}{Erd\H{o}s--R\'{e}nyi}
\newcommand{\mat}[1]{\mathsf{#1}}
\newcommand{\order}{\mathcal{O}}
\newcommand{\diffm}[3]{\frac{\mathrm{d}^{#1} #2}{\mathrm{d}#3^{#1}}}

\newcommand{\Ro}{\mathcal{R}_0}
\title{Distribution of outbreak sizes for SIR disease in finite populations}
\author{Joel C. Miller\footnote{La Trobe University}}
\date{\today}

\begin{document}

\maketitle

\begin{abstract}
  We consider the spread of a Susceptible-Infected-Recovered (SIR)
  disease through finite populations and derive an expression for the
  final size distribution.  Our derivation allows arbitrary
  distributions of the number of transmissions caused by an infected
  individual.  We show how this calculation can be used to infer
  parameters of the infectious disease through observations in
  multiple small populations.  The inference suffers from some
  identifiability difficulties, and it requires many observations to
  distinguish between parameter combinations that correspond to the
  same reproductive number.
\end{abstract}
\section{Introduction}
The spread of infectious disease remains a major source of morbidity and mortality.  Many important diseases are of "SIR" type: individuals begin susceptible, become infected through interactions with infected individuals, and when they recover they gain immunity to reinfection.  Examples include pandemic or seasonal influenza~\cite{miller2009signature,chowell2008seasonal}, Ebola~\cite{lewnard2014dynamics}, and Measles~\cite{grenfell2001travelling}.  Control of these diseases can be greatly facilitated by understanding the details of the individual-level stochasticity observed in transmission~\cite{lloyd2005superspreading}.  Even when they account for stochasticity at the individual level, most mathematical models of infectious disease spread are built in the infinite population limit~\cite{lloyd2005superspreading,miller:pgf}, and for understanding how disease spreads in finite populations, researchers usually turn to computer simulations of varying levels of complexity~\cite{bershteyn2018implementation,mniszewski2008episims,germann2006mitigation}.

In this paper we consider a fully mixed population of $N$ initially susceptible individuals.  We assume that the number of transmissions caused by an individual is given by some \emph{a priori} known distribution (the ``offspring distribution''), and that the recipient of each transmission is chosen uniformly at random from the rest of the population (with replacement, so $u$ might transmit twice to $v$, but $u$ will not transmit to itself).

We focus on two quantities:
\begin{itemize}
\item Firstly, we are interested in the probability $q_k$ the
disease infects exactly $k$ individuals.  
We show that the calculation of $q_k$ for all $k$ reduces to solving a linear system of the form $\mat{C} \vec{q} = \vec{1}$ where $\mat{C}$ is a lower-triangular matrix.  Because this system can be solved efficiently, we are able to use our results to infer parameters of the offspring distribution based on observed outbreak sizes.
\item Secondly, we are interested in calculating the probability that
  there are $m$ infections in a given disease generation conditional
  on the number of individuals of each status in the previous
  generation.  This calculation is more involved, but we are able to
  use it to infer more properties of a disease from our observations.
\end{itemize}
Once we find these, we show how they can be used to infer disease parameters from observations of outbreak final sizes in a number of populations.  We then extend this to show how observing individual generations improves our inference.

Our major result is the following:
{
\it
\begin{quote}
    Consider an SIR disease spreading in a finite population of $N$ individuals, and assume the
    number of transmissions $\ell$ an infected person will cause has a known
    distribution, with probability given by $p_\ell$.  Define $\mu(x)=
    \sum p_\ell x^\ell$.  Then, setting $q_k$ to be the probability that a single introduced infection would result in a total of exactly $k$ infections (including itself) for $k=1, \ldots, N$, we have
    \begin{align*}
    1 &= c_{1,1} q_1 \\
    1 &= c_{1,2}q_1 + c_{2,2}q_2 \\
    & \:\: \vdots \\
    1 &= c_{1,N}q_1 + c_{2,N} q_2 + \cdots + c_{N,N}q_N
    \end{align*}
    where 
    \[
    c_{k,M} = \left[ \mu\left(\frac{M-1}{N-1}\right)\right]^{-k} \prod_{j=1}^{k-1} \frac{M-j}{N-j} \, .
    \]
\end{quote}
}
The function $\mu(x)$ is the \emph{Probability Generating Function} (PGF) of the distribution.  This linear system of equations is triangular, so it can be solved quickly.  It is worth noting that if we start with  $c_{1,M}= 1/\mu\left(\frac{M-1}{N-1}\right)$, then for $k > 1$,
\[
c_{k,M} =   \frac{M-k+1}{N-k+1} \left[\mu \left( \frac{M-1}{N-1}\right)\right]^{-1}c_{k-1,M}
\]
which yields a rapid calculation of the coefficients.
If self-transmissions are allowed, then the only change is that
$(M-1)/(N-1)$ in the argument of $\mu$ is replaced by $M/N$.

This is a consequence of the following equivalent theorem expressed in the language of graph theory:
{\it 
\begin{quote}
 Consider a directed $N$-node multigraph created as follows: Each node $u$ has its out-degree assigned from a distribution with PGF $\mu(x)$.  For each edge coming out of $u$, the neighbor $v \neq u$ is chosen uniformly at random from the other $N-1$ nodes, with replacement (so edges may be repeated).
 
 Then setting $q_k$ to be the probability that a randomly chosen node has an out-component consisting of exactly $k$ nodes (including itself) for $k=1, \ldots, N$, we have
    \begin{align*}
    1 &= c_{1,1} q_1 \\
    1 &= c_{1,2}q_1 + c_{2,2}q_2 \\
    & \:\: \vdots \\
    1 &= c_{1,N}q_1 + c_{2,N} q_2 + \cdots + c_{N,N}q_N
    \end{align*}
    where 
    \[
    c_{k,M} = \left[ \mu\left(\frac{M-1}{N-1}\right)\right]^{-k} \prod_{j=1}^{k-1} \frac{M-j}{N-j} \, .
    \]
\end{quote}
}
The network class generated by assigning directed edges in this manner is related to the $k$-out networks~\cite{frieze2016introduction} as well as the inhomogeneous $k$-out networks~\cite{eletreby2018connectivity}.  It is distinct from these due to its directionality and the fact that it allows a node to transmit to the same target multiple times.

A small generalization of this allows for multiple introductions,
  still yielding a triangular linear system.  
For this generalization, we set $\chi(x)$ to be the PGF for the total
number of transmissions from outside (which could go to individuals
that are already infected).  Then the only change in the calculation
is that
\[
c_{k,M} = \frac{1}{\chi\left(\frac{M}{N}\right)} \left[\mu\left(\frac{M-1}{N-1}\right)\right]^{-k}
\prod_{j=0}^{k-1} \frac{M-j}{N-j}
\]
where $\chi(M/N)$ appears in the denominator and the product starts
from $j=0$ rather than $j=1$.

In some real-world applications, we may want to calculate the probability of a particular number infected in a given ``generation''.  Given the number susceptible $s_g$ and infected $i_g$ at generation $g$, the probability of $m$ infections at generation $g+1$ is given by
\[
P(i_{g+1}=m|s_g, i_g, N) = \sum_{\ell=m}^\infty p_{s_g}(\ell, m) \left(\frac{s_g}{N}\right)^{\ell} \frac{1}{\ell!} \left.\diffm{\ell}{}{x} [\mu(x)]^{i_g} \right|_{x=\frac{i_g+r_g}{N}}
\]

To derive these results, we first introduce some background theory
showing how the infectious disease problem is equivalent to the graph
theory question. Then we consider a simple case,
where with probability $p$ an infected individual transmits
independently to any given other individual at least once.  This
results in a binomial distribution of the number of individuals
receiving at least one transmission, with parameters $N-1$ and $p$.
We then move on to the more complicated cases where other
distributions are used.  For our final result, we derive the
probability that $m$ infections occur in generation $g+1$ given that a
population of $N$ individuals
there are $s_g$ susceptible and $i_g$ infected individuals at
generation $g$.  We end the paper by applying our results to the
problem of inferring parameter values based on simulated outbreaks.
We find that using outbreak data to distinguish between disease
parameters that correspond to the same basic reproductive number is difficult.  Although this appears to be a weakness of the approach, it also suggests that we can reasonably predict the possible outcomes of epidemics using just the basic reproductive number.

\section{Preliminaries}
\subsection{SIR disease and directed networks}
Before we build up the basic theory linking disease spread with
directed networks, we clarify our definition of a ``transmission''.
Once individual $u$ is infected, it transmits to others in the population.  When a transmission occurs it can be to any other 
individual $v$ regardless of $v$'s status and regardless of whether $u$ has transmitted to $v$ previously.  If that transmission occurs while $v$ is susceptible, then $v$ becomes infected.  If however $v$ is no longer susceptible, then the transmission has no effect.  So to be clear, a ``transmission'' does not require that the recipient be susceptible and become infected.

There is a mapping between the spread of an SIR disease and directed
networks if we make a few standard assumptions about the
disease~\cite{kiss:EoN, kenah:EPN}.  In particular, we assume that the
time at which individual $u$ becomes infected does not affect who
$u$ will transmit to.  So the probability that $u$ transmits to a
given set of nodes is the same whether $u$ is infected at the
beginning of the outbreak, the end, or any intermediate time.

Given this assumption, we can take a fatalistic view of the
transmission process.  Namely, that the individuals who would
receive transmission from $u$ (if $u$ ever becomes infected) are
chosen before the disease is introduced.  This defines a directed
network $G$.  Node $u$ has an edge to each node $v$ if and only if individual $u$ would transmit to individual $v$ at least once.  Once $G$ is defined and the
initial infection(s) chosen, the individuals that eventually become infected
are those which correspond to the nodes that are reachable from the initial infection(s) by following a path in $G$.  In other words, the
infected individuals are exactly those nodes in the out-component of
the initial infection (including the initial infection).

In our case, we have a finite set of $N$ individuals, and for each
individual the number of transmissions caused $\ell$ is chosen from a
given distribution [having probability generating function $\mu(x) =
\sum_\ell p_\ell x^\ell$].  For a given node $u$, once $\ell$ is
chosen we choose the recipients from the other $N-1$ nodes in the
population, with replacement.  So a node may send multiple
transmissions to the same target.  This builds $G$.  

We focus on determining the size distribution of the reachable set
given a random initial infection.  In the context of the random network
class, this is the size distribution of the out-component of a random node.

\begin{figure}
\begin{center}
\colorbox[gray]{0.95}{
\begin{minipage}{0.85\textwidth}\begin{algorithmic}
\Require \\
\texttt{Population} The individuals of the population\\
\texttt{OffspringDistribution} A function that chooses a
random number from the offspring distribution.
\Ensure \\ $G$: A directed network representing the potential transmissions.\\

\Function{GenerateNetwork}{Population, OffspringDistribution}
  \State $G$ $\gets$ Edgeless Directed MultiGraph with nodes from Population
  \For{$u$ in Population}
    \State OffspringCount $\gets$ OffspringDistribution()
    \For{counter in range(OffspringCount)}
      \State $v$ = RandomChoice(Population $\backslash$ \{$u$\})
      \State $G$.AddEdge($u$,$v$)
    \EndFor
  \EndFor
  \Return $G$
\EndFunction
\end{algorithmic}\end{minipage}}
\end{center}
\caption{Algorithm for generating a directed network from a population and known offspring distribution, assuming that each transmission goes to a randomly chosen individual from the population.  The input \texttt{Population} is the population.  \texttt{OffspringDistribution} is a function that returns a random value chosen from the offspring distribution.  For each individual $u$, we choose the individuals $v$ that it would transmit to if ever infected, and add those edges to the graph (psuedocode is modelled on networkx v2.2).}
\label{fig:algorithm}
\end{figure}

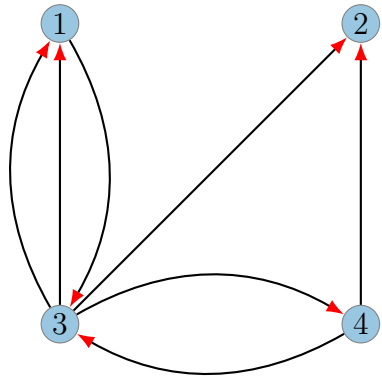
\begin{figure}
\centering
\begin{tikzpicture}
\node [BasicNode]  (1) at (0,0) {$1$};
\node [BasicNode]  (2) at (4,0) {$2$};
\node [BasicNode]  (3) at (0,-4) {$3$};
\node [BasicNode]  (4) at (4,-4) {$4$};
\path [-{Latex[red]}, thick, bend left, below, sloped] (1) edge node {} (3);
\path [-{Latex[red]}, thick, bend left,above, sloped] (3) edge node {} (1);
\path [-{Latex[red]}, thick, above, sloped] (3) edge node {} (2);
\path [-{Latex[red]}, thick, bend left, sloped, below] (3) edge node {} (4);
\path [-{Latex[red]}, thick, sloped, below] (4) edge node {} (2);
\path [-{Latex[red]}, thick, sloped, below] (3) edge node {} (1);
\path [-{Latex[red]}, thick, bend left, below, sloped] (4) edge node {} (3);
\end{tikzpicture}
\caption{A sample directed network generated by the algorithm of Fig.~\ref{fig:algorithm} (c.f., Fig. 6.14 of~\cite{kiss:EoN}).  If node $1$ is ever infected, it will transmit to node $3$.  If node $2$ is ever infected it will not transmit at all.  If node $3$ is ever infected, it will transmit twice to node $1$ and once each to nodes $2$ and $4$.  If node $4$ is ever infected, it will transmit to both $2$ and $3$.  Note that because we assume an SIR disease, a node is only infected the first time it receives a transmission.}
    \label{fig:my_label}
\end{figure}

\section{The simplest case}
\label{sec:simple}
We start with a simpler case in which each individual produces a Poisson-distributed number of transmissions with mean $\Ro$.  Each time an individual $u$ transmits, the recipient is selected from the other $N-1$ individuals in the population (with replacement).

We note that when the number of transmissions is Poisson-distributed, whether one individual receives at least one transmission is independent of what happens to others.  This property will simplify our analysis here.  It does not hold for other distributions, so our more general derivation is more difficult.

The number of individuals receiving at least one transmission from a
given infected individual $u$ is binomially-distributed with each of
the $N-1$ other individuals chosen independently with probability $p = 1- e^{-\Ro/(N-1)}$.

We can assume without loss of generality that the population is numbered
$1, \ldots, N$ and the infection is introduced in individual $1$.
Infection will spread to the out-component of node $1$ in the
corresponding directed network.
We define $q_k$ to be the probability that the out-component of node
$1$ has exactly $k$ nodes, including node $1$.  Equivalently, this is
the probability that the initial infection results in a total of $k$ infections.

We briefly outline our strategy to calculate $q_k$.  We first note that the probability that nodes
$1, \ldots, M$ in
the directed network $G$ representing the potential transmissions have no edges to any node in $M+1,
\ldots, N$ is $(1-p)^{M(N-M)}$.  Then we will find another
expression for this same probability that arises as a summation with each term
depending on $q_k$ for $k = 1, \ldots, M$.  When we perform this for $M=1, 2, \ldots, N$, we
arrive at a system of equations of the form $\sum_{k=1}^M d_{k,M} q_k
= (1-p)^{M(N-M)}$.  This can be represented by a triangular
matrix and solved quickly.

.




\subsection{Equations}
\label{sec:equations}
We now fill in the details of our derivation.  
We take $M$ and $p$ as given.  We can find the probability that none of
$1, \ldots, M$ has an edge to any of $M+1, \ldots, N$ by
noting that there are $M(N-M)$ pairs where the first is chosen from
$1, \ldots, M$ and the second is chosen from $M+1, \ldots, N$.  The
independent probability for each directed pair of having no edge is $1-p$.  So
the probability none of the edges exist is $(1-p)^{M(N-M)}$ (note that edges in the opposite direction are allowed).

We now look for another calculation of this probability.  We make an
observation that 
{
\it 
\begin{quote}
None of $1, \ldots,
M$ have edges to any of $M+1, \ldots, N$ if and only if
\begin{itemize}
\item All nodes in the out-component of node $1$ in the directed graph
  are in $1, \ldots, M$. and
\item taking $k$ to be the size of the out-component of node $1$
  (including node $1$), the other $M-k$ nodes in $1, \ldots, M$ have
  no edges to any of $M+1, \ldots, N$.
\end{itemize}
\end{quote}
}
\noindent This observation is shown in Fig.~\ref{fig:both_derivations}, and a
straightforward proof is provided in the appendix.

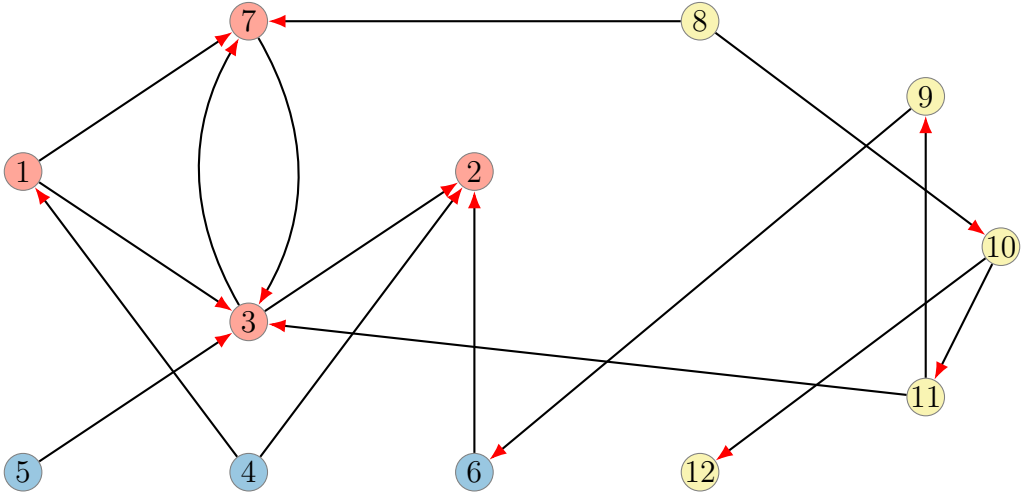
\begin{figure}
\centering
\begin{tikzpicture}
\node [BasicNode, fill=colorI!40]  (1) at (0,0) {$1$};
\node [BasicNode, fill=colorI!40]  (7) at (3,2) {$7$};
\node [BasicNode, fill=colorI!40]  (3) at (3,-2) {$3$};
\node [BasicNode, fill=colorI!40] (2) at (6, 0) {$2$};
\node [BasicNode]  (4) at (3,-4) {$4$};
\node [BasicNode]  (5) at (0,-4) {$5$};
\node [BasicNode]  (6) at (6,-4) {$6$};

\node [BasicNode, fill = colorf!40]  (8) at (9,2) {$8$};
\node [BasicNode, fill = colorf!40]  (9) at (12,1) {$9$};
\node [BasicNode, fill = colorf!40]  (10) at (13,-1) {$10$};
\node [BasicNode, fill = colorf!40]  (11) at (12,-3) {$11$};
\node [BasicNode, fill = colorf!40]  (12) at (9,-4) {$12$};

\path [-{Latex[red]}, thick, above, sloped] (1) edge node {} (3);
\path [-{Latex[red]}, thick, above, sloped] (1) edge node {} (7);
\path [-{Latex[red]}, thick, above, sloped] (3) edge node {} (2);
\path [-{Latex[red]}, thick, bend left, below, sloped] (3) edge node {} (7);
\path [-{Latex[red]}, thick, bend left, above, sloped] (7) edge node {} (3);
\path [-{Latex[red]}, thick, sloped, below] (5) edge node {} (3);
\path [-{Latex[red]}, thick, below, sloped] (4) edge node {} (1);
\path [-{Latex[red]}, thick, below, sloped] (4) edge node {} (2);
\path [-{Latex[red]}, thick, below, sloped] (6) edge node {} (2);

\path [-{Latex[red]}, thick, below, sloped] (8) edge node {} (10);
\path [-{Latex[red]}, thick, below, sloped] (8) edge node {} (7);
\path [-{Latex[red]}, thick, below, sloped] (9) edge node {} (6);
\path [-{Latex[red]}, thick, below, sloped] (10) edge node {} (12);
\path [-{Latex[red]}, thick, below, sloped] (10) edge node {} (11);
\path [-{Latex[red]}, thick, below, sloped] (11) edge node {} (9);
\path [-{Latex[red]}, thick, below, sloped] (11) edge node {} (3);
\end{tikzpicture}
\caption{An illustration of the observation of Section~\ref{sec:equations}:  
  Because the out-component of $1$ lies entirely within $1, \ldots, 7$
  (and therefore none of those nodes has edges to $8, \ldots, 12$)
  and none of the other nodes in $1, \ldots, 7$ have edges to $8, \ldots, 12$,
  we know there are no edges from any node in $1, \ldots, 7$ to any
  node in $8,\ldots, 12$.  The converse also holds: because there are no
  edges from $1, \ldots, 7$ to $8, \ldots, 12$, we can be certain that
  the outcomponent of $1$ lies entirely within $1, \ldots, 7$ and all
  other nodes in $1, \ldots, 7$ have no edges to $8, \ldots, 12$.}
\label{fig:both_derivations}
\end{figure}

So our alternate calculation of the probability of no edges from the
first $M$ nodes to nodes $M+1, \ldots, N$ comes from summing up over
all $k$ the probability node $1$ has an out-component of $k$ nodes,
all of which are within $1, \ldots, M$ and there are no edges from any
of the other $M-k$ still-susceptible nodes to any of the nodes in
$M+1, \ldots, N$.  

To calculate this, we take $k$ as given.  We have
\begin{itemize}
\item The probability that the out-component of node $1$ is made up of
exactly $k$ nodes (including node $1$) is by definition $q_k$.
\item Given that the out-component of $1$ has $k$ nodes, the
  probability that the  $k-1$ nodes reachable from $1$ (not including $1$) are
  in $2, \ldots, M$ is $\binom{M-1}{k-1}/\binom{N-1}{k-1}$.
\item Given that the $k$ nodes in the out-component of $1$ are all in $1, \ldots, M$,
  the probability that the other $M-k$ nodes in $1, \ldots, M$ also do
  not have edges to any of $M+1, \ldots, N$ is $(1-p)^{(M-k)(N-M)}$.
\end{itemize}
Thus the probability that the out-component has $k$ nodes, those nodes
are entirely within $1, \ldots, M$, and none of the other $M-k$ nodes
in $1, \ldots, M$ have edges to nodes in $M+1, \ldots, N$ is the
product 
\[
q_k \frac{\binom{M-1}{k-1}}{\binom{N-1}{k-1}}(1-p)^{(M-k)(N-M)}
\]
Summing over all $k$ gives the probability that the out-component of
$1$ lies within $1, \ldots, M$ and none of the other nodes in $1,
\ldots, M$ have edges to $M+1, \ldots, N$.  We have observed above
that this is exactly the probability that there are no edges from $1,
\ldots, M$ to $M+1, \ldots, N$.  So
\[
(1-p)^{M(N-M)}  = \sum_{k=1}^M q_k \frac{\binom{M-1}{k-1}}{\binom{N-1}{k-1}}(1-p)^{(M-k)(N-M)}
\]
This can be rewritten
\[
1  = \sum_{k=1}^M c_{k,M} q_k 
\]
where
\begin{align}
c_{k,M} &=
\frac{(1-p)^{(M-k)(N-M)}}{(1-p)^{M(N-M)}}\frac{\binom{M-1}{k-1}}{\binom{N-1}{k-1}}\nonumber\\
&=  (1-p)^{-k(N-M)} \prod_{j=1}^{k-1} \frac{M-j}{N-j}\label{eqn:ckm}
\end{align}


Performing this sum for every $M$, we get the system
\begin{align*}
1 &= c_{1,1} q_1 \\
1 &= c_{1,2}q_1 + c_{2,2}q_2 \\
& \:\: \vdots \\
1 &= c_{1,N}q_1 + c_{2,N} q_2 + \cdots + c_{N,N}q_N
\end{align*}
We can interpret $c_{k,M}q_M$ as the probability that the
out-component has $k$ nodes given that there are no edges from
$1, \ldots, M$ to $M+1, \ldots, N$.  The matrix of coefficients is
lower triangular, and so the numerical solution of this system is
efficient, once we determine the coefficients.

Notice that in Equation~\eqref{eqn:ckm} the disease properties only
appear in the term $(1-p)^{-k(N-M)}$.  Reviewing this term, it comes
from the probability that the $M-k$ individuals in $1, \ldots, M$ who
remain uninfected would not transmit to any individual in $M+1,
\ldots, N$ divided by the probability that individuals $1, \ldots, M$
have no transmissions to $M+1, \ldots, N$.  This is the only disease-dependant term, and when we investigate other distributions of numbers of transmissions it is the only term that is modified.

Our result here matches the distribution for the component size of a
chosen node in a finite (undirected) \erdosrenyi{} network found by~\cite{rath2018moment}.  When transmission probabilities are symmetric and transmissions occur independently, the directed network representing transmissions can be replaced by an undirected network~\cite{newman2002spread,kiss:EoN,kenah:EPN,meyers:contact,meyers:directed,kenah:second,miller:heterogeneity,hastings:series, durrett2007random}.  In this case each edge would exist independently with probability $p$.  So we would expect the same size distribution as seen in undirexted \erdosrenyi{} networks.

\section{General offspring distributions}
It is frequently observed that some individuals cause significantly more infections than others~\cite{lloyd2005superspreading,meyers:sars,shen2004superspreading}.  The offspring distribution is decidedly not Poissonian.

The Poisson offspring distribution assumed in the derivation of Eq.~\eqref{eqn:ckm} emerges from a stochastic model in which all infections have the same duration and infected individuals transmit with the same constant rate.  One modification of this assumption allows the infectious period to have an arbitrary distribution but infected individuals still transmit at the same constant rate.  Using this, a similar result to Eq.~\eqref{eqn:ckm} was derived by~\cite{ball1986unified}.  The proof is similar to what we used above.

In this section, we fully generalize the result to arbitrary offspring distributions.  We take an arbitrary (known) distribution of the number of transmissions, and define
\[
\mu(x) = \sum_\ell p_\ell x^{\ell}
\]
where $p_\ell$ is the probability a random individual $u$ causes $\ell$ transmissions.  Each transmission from individual $u$ goes to a randomly chosen individual (other than $u$), possibly the same as a previous transmission from $u$.

We will first study the case in which infection is introduced a single time, and then consider modifications allowing us to explore an arbitrary number of introductions.  

\subsection{Single Introduction}

We follow our previous argument.  As before, we choose some $M$ with $1 \leq M \leq N$.  The probability that the recipient $v$ of a given transmission from $u \in \{1,
\ldots, M\}$ is also in $\{1, \ldots, M\}$  is $(M-1)/(N-1)$ (the $-1$s
appear because we exclude self-transmissions).  The probability that
all recipients of transmissions from $u$ are restricted to be within the first $M$
individuals is thus
\[
\sum_\ell p_\ell \left(\frac{M-1}{N-1}\right)^\ell = \mu\left(\frac{M-1}{N-1}\right)
\]

The probability that there are no
edges from $1, \ldots, M$ to any node in $M+1,
\ldots, N$ is $\left[\mu\left(\frac{M-1}{N-1}\right)\right]^{M}$.  As before we find another
expression for this as a sum depending on $q_k$.
\begin{itemize}
    \item The probability that the out-component of node $1$ is made up of exactly $k$ nodes (including node $1$) is by definition $q_k$.
\item Given that the out-component of $1$ has $k$ nodes, the
  probability that the  $k-1$ nodes reachable from $1$ (not including $1$) are
  in $2, \ldots, M$ is $\binom{M-1}{k-1}/\binom{N-1}{k-1}$.
    \item Given that the $k$ nodes in the out-component of $1$ are all in $1, \ldots, M$, the probability that the other $M-k$ nodes in $1, \ldots, M$ also have no edges to any of $M+1, \ldots, N$ is $[\mu((M-1)/(N-1))]^{(M-k)}$.
\end{itemize}
So the probability node $1$ has an out-component of size exactly $k$
contained entirely within $1, \ldots, M$  and there are
no other edges from $1, \ldots, M$ to $M+1, \ldots, N$ is $q_k
[\mu((M-1)/(N-1))]^{M-k} \prod_{j=1}^{k-1} (M-j)/(N-j)$.  Summing over
all $k$ we find that the probability of no edges from $1, \ldots, M$
to $M+1, \ldots, N$ is
\[
\left[\mu\left(\frac{M-1}{N-1}\right)\right]^{M} = \sum_{k =1}^M q_k\left[\mu\left(\frac{M-1}{N-1}\right)\right]^{(M-k)} \frac{\binom{M-1}{k-1}}{\binom{N-1}{k-1}}
\]
As before, this becomes
\[
1 = \sum_k c_{k,M} q_k
\]
where we now have
\begin{equation}
c_{k,M} = \left[\mu\left(\frac{M-1}{N-1}\right)\right]^{-k} \prod_{j=1}^{k-1} \frac{M-j}{N-j}
\label{eqn:genericckm}
\end{equation}
playing the role of Eq.~\eqref{eqn:ckm}.


The $(1-p)^{-k(N-M)}$ in Equation~\eqref{eqn:ckm} is thus replaced by $[\mu((M-1)/(N-1))]^{-k}$.  The other parts of Equation~\eqref{eqn:ckm} remain the same.  Putting this all together, we again find 
\begin{align*}
1 &= c_{1,1} q_1 \\
1 &= c_{1,2}q_1 + c_{2,2}q_2 \\
& \:\: \vdots \\
1 &= c_{1,N}q_1 + c_{2,N} q_2 + \cdots + c_{N,N}q_N
\end{align*}
where here $c_{k,M} =  \left[\mu\left(\frac{M-1}{N-1}\right)\right]^{-k} \prod_{j=1}^{k-1} \frac{M-j}{N-j}$.

\subsubsection{Special case of a Poisson distribution}
Our model in Section~\ref{sec:simple} corresponds to assuming a Poisson distribution with mean $\Ro$. We show that this arises as a special case of our result here.  For the Poisson distribution, we have
\[
\mu(x) = e^{-\Ro (1-x)}
\]
We found $p=1- e^{-\Ro /(N-1)}$ and so
\[
\mu(x)=
[(1-p)^{N-1}]^{1-x} = (1-p)^{(N-1)(1-x)}
\]
Finally,
\begin{align*}
    \left[\mu\left(\frac{M-1}{N-1}\right)\right]^{-k} &= \left[(1-p)^{(N-1)\left(1-\frac{M-1}{N-1}\right)}\right]^{-k}\\
    &= (1-p)^{-k(N-1-[M-1])}\\
    &= (1-p)^{-k(N-M)}
\end{align*}
which yields our result in Section~\ref{sec:simple}.

\subsection{Multiple Introductions}

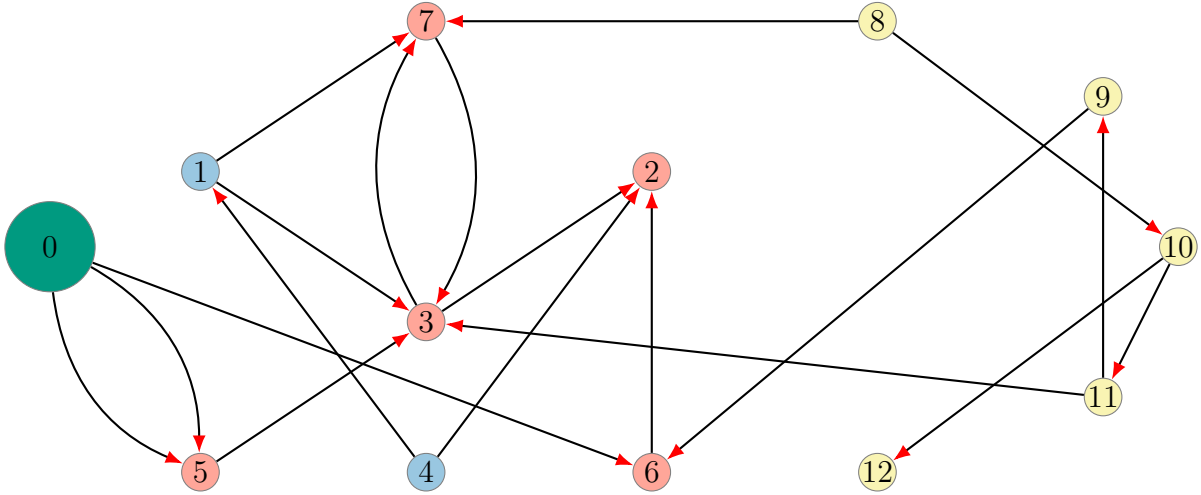
\begin{figure}
\centering
\begin{tikzpicture}
\node [BasicNode, fill=colorc, minimum size=12mm] (0) at (-2,-1) {$0$};
\node [BasicNode]  (1) at (0,0) {$1$};
\node [BasicNode, fill=colorI!40]  (7) at (3,2) {$7$};
\node [BasicNode, fill=colorI!40]  (3) at (3,-2) {$3$};
\node [BasicNode, fill=colorI!40] (2) at (6, 0) {$2$};
\node [BasicNode]  (4) at (3,-4) {$4$};
\node [BasicNode, fill=colorI!40]  (5) at (0,-4) {$5$};
\node [BasicNode, fill=colorI!40]  (6) at (6,-4) {$6$};

\node [BasicNode, fill = colorf!40]  (8) at (9,2) {$8$};
\node [BasicNode, fill = colorf!40]  (9) at (12,1) {$9$};
\node [BasicNode, fill = colorf!40]  (10) at (13,-1) {$10$};
\node [BasicNode, fill = colorf!40]  (11) at (12,-3) {$11$};
\node [BasicNode, fill = colorf!40]  (12) at (9,-4) {$12$};

\path [-{Latex[red]}, thick, above, sloped] (0) edge node {} (6);
\path [-{Latex[red]}, thick, above, sloped, bend left] (0) edge node {} (5); 
\path [-{Latex[red]}, thick, above, sloped, bend right] (0) edge node {} (5);

\path [-{Latex[red]}, thick, above, sloped] (1) edge node {} (3);
\path [-{Latex[red]}, thick, above, sloped] (1) edge node {} (7);
\path [-{Latex[red]}, thick, above, sloped] (3) edge node {} (2);
\path [-{Latex[red]}, thick, bend left, below, sloped] (3) edge node {} (7);
\path [-{Latex[red]}, thick, bend left, above, sloped] (7) edge node {} (3);
\path [-{Latex[red]}, thick, sloped, below] (5) edge node {} (3);
\path [-{Latex[red]}, thick, below, sloped] (4) edge node {} (1);
\path [-{Latex[red]}, thick, below, sloped] (4) edge node {} (2);
\path [-{Latex[red]}, thick, below, sloped] (6) edge node {} (2);

\path [-{Latex[red]}, thick, below, sloped] (8) edge node {} (10);
\path [-{Latex[red]}, thick, below, sloped] (8) edge node {} (7);
\path [-{Latex[red]}, thick, below, sloped] (9) edge node {} (6);
\path [-{Latex[red]}, thick, below, sloped] (10) edge node {} (12);
\path [-{Latex[red]}, thick, below, sloped] (10) edge node {} (11);
\path [-{Latex[red]}, thick, below, sloped] (11) edge node {} (9);
\path [-{Latex[red]}, thick, below, sloped] (11) edge node {} (3);
\end{tikzpicture}
\caption{Illustration of the outcome for multiple introductions (in this case through two outside transmissions to individual $5$ and one to individual $6$), assuming the same internal transmissions as in Fig.~\ref{fig:my_label}.  Note that the auxiliary node $0$ receives no transmissions.  The fact that the nodes reachable from $0$ lie within $1, \ldots, 7$ and the other nodes in $1, \ldots, 7$ have no edges to $8, \ldots, 12$ is equivalent to the fact that there are no edges from $0, \ldots, 7$ to $8, \ldots, 12$.}
\label{fig:multiIntro}
\end{figure}

We are interested in the outcome of multiple introductions, where the introduction is done as multiple transmissions from outside the small population.  We assume that the number of transmissions from outside is a random variable, and that the transmission goes to a randomly chosen individual (with replacement --- that is, the same individual may receive multiple transmissions). We assume that the PGF for the number of introductions is $\chi(x)$. 
 The results will immediately carry over to a fixed number of introductions (choosing the initial infections with replacement), though some modifications will be needed if the introductions are done without replacement.

To do this, we recast the new problem to mimic the original.  We add an auxiliary individual $0$, which will be
the source of introductions.  Individual $0$ causes a number of transmissions with PGF
$\chi(x)$ and the transmissions can go to any of the $N$ nodes in
$1, \ldots, N$ with replacement.  The other infected nodes each cause a number of transmissions with
PGF $\mu(x)$ and a given node $u\neq0$ can transmit to any node in
$1, \ldots, N$ except itself.  Note that no transmissions go to the
auxiliary node $0$; it has in-degree $0$.

We first calculate the probability that the recipients of all of the
potential transmissions from $0$ are restricted to $1,
\ldots, M$.  Each transmission has probability $M/N$ of reaching one
of these nodes.  So summing over all $\ell$ transmissions, the
probability they are all to nodes in $1, \ldots, M$ is
$\chi(M/N)$.  For the nodes in $1, \ldots, M$, the probability that
all recipients of their transmissions are in $1, \ldots, M$ is 
$[\mu((M-1)/(N-1)]^M$.  Combining these, the probability of no
transmissions from $0, \ldots, M$ to $M+1, \ldots, N$ is  $\chi(M/N) [\mu((M-1)/(N-1))]^M$. 

Now we calculate this same probability through $q_k$.  
\begin{itemize}
\item The probability that the out-component of node $0$ has exactly
  $k$ nodes \emph{not} including node $0$ is defined to be $q_k$.
\item  If the out-component of node $0$ has exactly
  $k$ nodes  \emph{excluding} node $0$, the probability that those $k$ are in $1, \ldots, M$
and none are in $M+1, \ldots, N$ is $\binom{M}{k} / \binom{N}{k}$.
\item If the out-component of node $0$ has exactly $k$ other nodes, all within $1,
  \ldots, M$, the probability that the $M-k$ unreached nodes in $1,
  \ldots, M$ have no edges to
any nodes in $M+1, \ldots, N$ is $[\mu((M-1)/(N-1))]^{M-k}$. 
\end{itemize}
Putting this together, the probability that the out-component of $0$
has exactly $k$ nodes excluding $0$, they occur within $1, \ldots,
M$, and none of the unreached nodes in $1, \ldots, M$ has an edge to
any node in $M+1, \ldots, N$ is
\[
q_k \left[\mu\left(\frac{M-1}{N-1}\right)\right]^{M-k}
\prod_{j=0}^{k-1} \frac{M-j}{N-j}
\]
Note that the product starts at $j=0$, not $j=1$.  Summing over all
$k$, we find that the probability that none of $0, \ldots, M$ has an
edge to any node in $M+1, \ldots, N$ is
\[
\sum_k q_k \left[\mu\left(\frac{M-1}{N-1}\right)\right]^{M-k}
\prod_{j=0}^{k-1} \frac{M-j}{N-j} \, .
\]
We get
\[
\chi\left(\frac{M}{N}\right) \left[\mu\left(\frac{M-1}{N-1}\right)\right]^M = \sum_k q_k \left[\mu\left(\frac{M-1}{N-1}\right)\right]^{M-k}
\prod_{j=0}^{k-1} \frac{M-j}{N-j} \, ,
\]
and so finally
\begin{align*}
1 &= c_{1,1} q_1 \\
1 &= c_{1,2}q_1 + c_{2,2}q_2 \\
& \:\: \vdots \\
1 &= c_{1,N}q_1 + c_{2,N} q_2 + \cdots + c_{N,N}q_N
\end{align*}
where 
\begin{equation}
c_{k,M} = \frac{1}{\chi\left(\frac{M}{N}\right)} \left[\mu\left(\frac{M-1}{N-1}\right)\right]^{-k}
\prod_{j=0}^{k-1} \frac{M-j}{N-j} \, .
\label{eqn:multickm}
\end{equation}

\subsubsection{Special case of a single introduction}
When there is a single introduced infection, $\chi(x) = x$, so
$\chi(M/N) = M/N$ appears in the denominator in the right hand side of
Eq.~\eqref{eqn:multickm}.  The $j=0$ term in the
product is also $M/N$.  These cancel one another.  The remaining expression is
identical to our earlier result in Eq.~\eqref{eqn:genericckm}.

\section{Temporal dynamics}
Let us assume that the infections can be clearly distinguished by generation.  In generation $g$ there are $i_g$ infections, $s_g$ susceptible individuals, and $r_g$ recovered, with $N=s_g+i_g+r_g$ constant.

Set $p_{k|i_g}$ to be the probability that $i_g$ individuals cause exactly $k$ transmissions.  Then $p_{k|i_g}$ is the coefficient of $x^k$ in $[\mu(x)]^{i_g}$.  Given $k$ transmissions, the probability that $\ell$ of them are to the $s_g$ susceptible individuals (possibly with repetition) is $\binom{k}{\ell} \left(\frac{s_g}{N}\right)^\ell \left(\frac{i_g+r_g}{N}\right)^{k-\ell}$.  So the probability of $\ell$ transmissions to susceptible individuals is 
\begin{align*}
\sum_{k} p_{k|i_g}\binom{k}{\ell} \left(\frac{s_g}{N}\right)^\ell \left(\frac{i_g+r_g}{N}\right)^{k-\ell}&= \left(\frac{s_g}{N}\right)^\ell \sum_k \frac{k!}{\ell! (k-\ell)!} p_{k|i_g} \left(\frac{i_g+r_g}{N}\right)^{k-\ell}\\
&= \left(\frac{s_g}{N}\right)^\ell \frac{1}{\ell!}\left.\diffm{\ell}{}{x} [\mu(x)]^{i_g} \right|_{x=\frac{i_g+r_g}{N}} 
\end{align*}

We take $m$ to denote the number of distinct susceptible individuals receiving transmissions.  The probability that $m$ new infections occur given $s_g$ susceptible and $i_g$ infected individuals is
\begin{equation}
P(i_{g+1}=m|s_g, i_g, N) = \sum_{\ell=m}^\infty p_{s_g}(\ell, m) \left(\frac{s_g}{N}\right)^{\ell} \frac{1}{\ell!} \left.\diffm{\ell}{}{x} [\mu(x)]^{i_g} \right|_{x=\frac{i_g+r_g}{N}}
\label{eqn:Pigp1}
\end{equation}
Where we define $p_{s}(m, \ell)$ to be the probability of $m$ distinct balls found when $\ell$ balls are chosen with replacement from a set of $s$ balls.  For $\ell\geq m >0$ it satisfies the relation 
\[
p_{s}(m,\ell) = \frac{(s-m+1)}{s}p_s(m-1,\ell-1) + \frac{m }{s}p_s(m, \ell-1)
\]   
Additionally we have $p_s(0,0)=1$.  If $\ell$ and $m$ do not satisfy either $\ell \geq m >0$ or $\ell=m=0$, then $p_s(m,\ell)=0$.

\section{Application to inference}
\label{sec:example}
In this section we will show how our results can be used to infer parameters of a disease spreading in a set of small communities.  We will simulate some outbreaks with known parameters and then attempt to infer those parameters.  The code used to perform the simulations and the inference is provided as a supplement.

\subsection{Parameter Inference from Final Size}
Consider a set of $s$ small communities, consisting of $N_1$, $N_2$, \ldots,
$N_s$ individuals each.  We assume that each community has exactly one
introduced infection, and that we observe both the size of the
outbreak in each community $k_i$ and the size of the community $N_i$.

We assume the probability distribution comes from a known family, but with some unknown parameters.  Our goal is to determine the parameters given our prior knowledge of the parameter values $P(\Theta)$ and the observed data ($k_i$ and $N_i$), which we represent by $X$.  We use Bayes' Theorem~\cite{hoff2009first}:
\begin{equation}
\label{eqn:bayes}
P(\Theta|X) = \frac{P(\Theta,X)}{P(X)} = \frac{P(X|\Theta) P(\Theta)}{P(X)}
\end{equation}
For each given $\Theta$, we find $P(X|\Theta)$ using the techniques
described above.  $P(X)$ is found by integrating all possible
$\Theta$ [or equivalently by normalizing the collection
$P(X|\Theta)P(\Theta)$].

We assume that the offspring distribution has a negative binomial distribution, parameterized by $r$ and $p$ (where $p$ is the probability of success for each trial and the integer $r$ is the number of failed trials before the process stops).  As there are multiple parameterizations of the negative binomial distribution, we note that for this distribution the PGF is 
\[
\mu(x) = \left[\frac{1-p}{1-px}\right]^r
\]
The mean of this distribution is $\Ro = pr/(1-p)$ and the variance is $pr/(1-p)^2$.

\begin{figure}
    \centering
    \includegraphics[width=\textwidth]{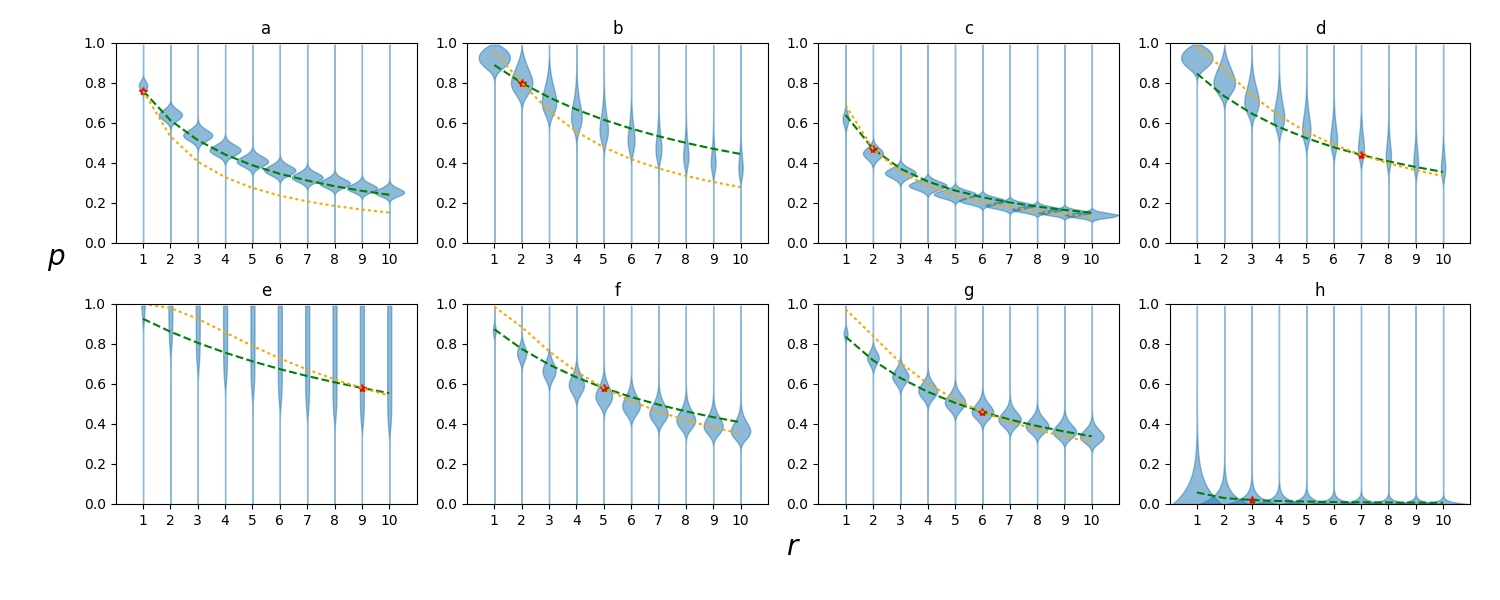}
      \caption{Scaled posterior probabilities for Table~\ref{table:outbreaks}.  The star denotes the true parameter values.  The shading denotes the inferred probability of a given $n$ and $p$, with the total shaded area in each figure corresponding to probability $1$ (note that $n$ is discrete while $p$ is continuous).  The dotted line shows parameters that would have the same large population limit epidemic probability as the true parameters.  The dashed line shows parameters with the same $\Ro$ as the true parameters (which would infect the same population proportion in an epidemic in the large population limit). We seem to be able to accurately infer $\Ro$, but it is difficult to learn much more.  \textbf{top}: posterior distribution from the data in the first four lines of the table.
      \textbf{bottom}: posterior distribution from the last four
      lines of the table.  Note that in (e), all possible infections occurred, and there is a large collection of parameter values for which this is likely, so there is little certainty about the parameter values.}
    \label{fig:infer}
\end{figure}

\begin{figure}
    \centering
    \includegraphics[width=\textwidth]{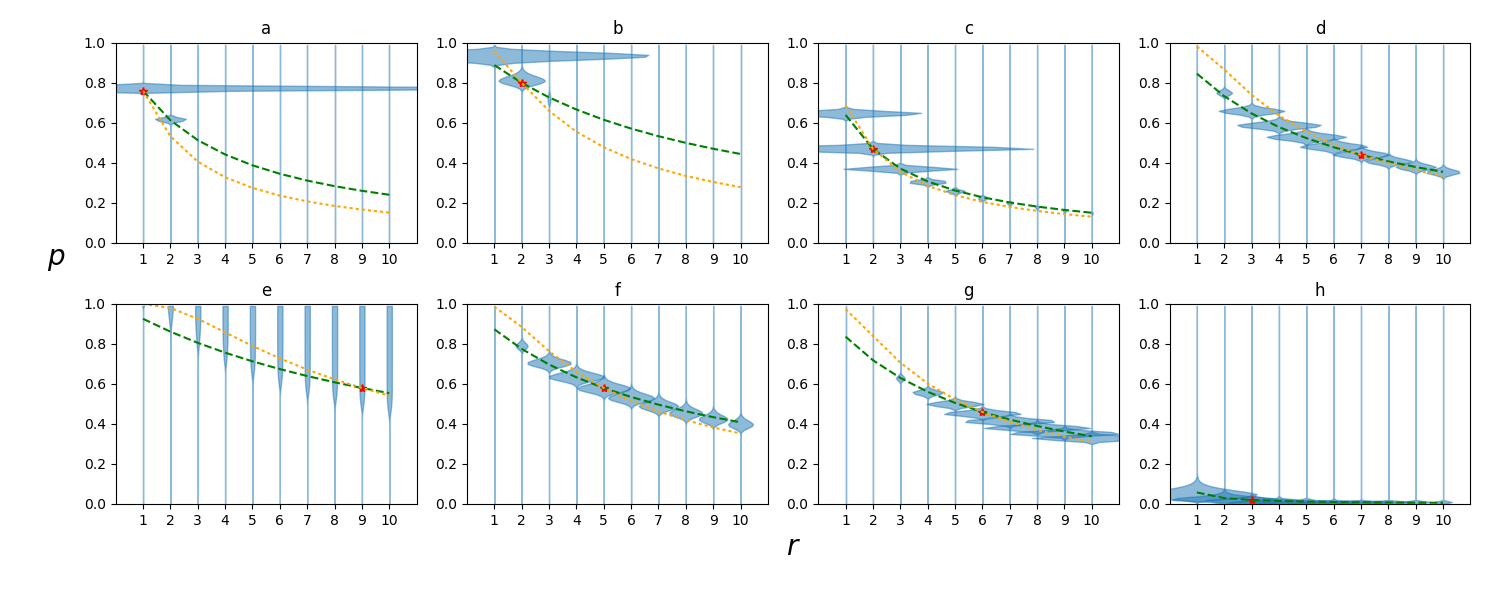}
      \caption{Scaled posterior probabilities for ten times as many observations in each population size as in Figure~\ref{fig:infer}.  The inference is able to improve the estimate.  However, it has difficulty distinguishing parameters with similar $\Ro$ or similar epidemic probabilities.  }
    \label{fig:infer_more}
\end{figure}

For our prior, we assume that $r$ is in $1,2,\ldots, 10$, each with equal probability, and $p$ is in $0, 0.01, 0.02, \ldots, 0.99$, again each with equal probability.  We choose one value of $p$ and $r$, and simulate outbreaks in 10 populations, one for each size in $5$, $10$, $15$, \ldots, $50$.  We then use the observations to infer $r$ and $p$.  We  repeat with another value for $r$ and $p$.  We achieve the following table:\footnote{Note that the random seeds leading to these data were chosen so that Fig.~\ref{fig:infer} would show a range of typical outcomes, and thus could plausibly contain some implicit biases.  Figures~\ref{fig:infer_more} and~\ref{fig:infer_dynamics} were not considered when choosing the seeds.}

\begin{table}[h]
\begin{center}
\begin{tabular}{c||cc||c@{\qquad}c@{\qquad}c@{\qquad}c@{\qquad}c@{\qquad}c@{\qquad}c@{\qquad}c@{\qquad}c@{\qquad}c}
\toprule 
\multicolumn{1}{c||}{} &
 \multicolumn{2}{c||}{Parameters} & \multicolumn{10}{c}{Number infected in simulated outbreaks in population of size:} \\
label &$r$ & $p$ & 5 & 10 & 15 & 20 & 25 & 30 &35& 40& 45 & 50 \\
\midrule
a& 1 & 0.76 & 5 & 10 & 15 & 18 & 25 & 26 & 34 & 1 & 44 & 49\\
b& 2 & 0.80 & 5 & 10 & 15 & 20 & 25 & 30 & 35 & 40 & 45 & 1\\
c&2 & 0.47 & 4 & 3 & 13 & 1 & 17 & 22 & 1 & 32 & 25 & 19\\
d&7 & 0.44 & 5 & 10 & 15 & 20 & 25 & 30 & 35 & 40 & 1 & 50\\
e&9 & 0.58 & 5 & 10 & 15 & 20 & 25 & 30 & 35 & 40 & 45 & 50\\
f&5 & 0.58 & 5 & 10 & 15 & 20 & 25 & 30 & 35 & 40 & 44 & 50\\
g&6 & 0.46 & 5 & 10 & 15 & 20 & 25 & 30 & 35 & 38 & 45 & 50\\
h&3 & 0.02 & 1 & 1 & 1 & 1 & 1 & 1 & 1 & 1 & 1 & 1
\end{tabular}
\end{center}
\caption{For each pair of parameter values, we perform 10 simulations, one in each population size and record the total number of infections.  This data is used in Fig.~\ref{fig:infer} to infer disease parameters.  Figure~\ref{fig:infer_more} uses these simulations plus nine more for each population size (thus a total of 100 simulations) to perform a more accurate inference, and Fig.~\ref{fig:infer_dynamics} uses just these ten simulations, but including the sizes of each generation.}
\label{table:outbreaks}
\end{table}

For each row of the table, Fig.~\ref{fig:infer} shows the a plot of the posterior probability distribution for the parameters.  In most cases, there is a relatively high probability assigned to the true parameter values, with a relatively thin region of plausible parameters.  In the $r=9$, \ $p=0.08$ case all possible infections occurred and it is difficult to distinguish between the most infectious cases where this is likely.  In the $r=3$, \ $p=0.02$ case, no additional infections occurred, and this is difficult to distinguish between the least infectious cases.  Interestingly for $r=5$, \ $p=0.58$, a single individual escaped in the $N=45$ population, which allows for reasonably good parameter estimation.

To help understand the structure of our observations, we first note that there are typically two types of outbreaks in a large population: non-epidemic outbreaks and epidemic outbreaks.  
\begin{itemize}
    \item In a non-epidemic outbreak, the disease is entirely self-limited: it dies out because the infected individuals fail to transmit, either because the average number of transmissions $\Ro$ is less than one or because the first few infected individuals failed to transmit further simply due to stochastic luck.  
    \item In an epidemic outbreak, the disease grows large enough that stochastic die-out does not occur.  Because it is an SIR disease, it eventually dies out because many of the new transmissions are going to individuals who are already infected. 
\end{itemize}
In a large population then, the probability of a non-outbreak epidemic is closely related to the offspring distribution, in particular the probability of no transmissions.  In fact the probability of no epidemic is given by the smallest solution to $x = \mu(x)$ in $[0,1]$~\cite{miller:pgf}.  

On the other hand, if there is a large outbreak then the central limit theorem comes into play.  It acts to obscure some of the properties we are trying to infer.  The number of transmissions that occur is well-approximated by the expected number per infected individual, $\Ro$, times the total number infected.  We can then predict the total number infected by calculating the expected number of successful infections to occur for a given number of transmissions.  These two conditions give a consistency relation that yields a ``final size relation''.  The only detail of the offspring distribution that goes into this relation is the average.  This underlies the sometimes-surprising consistency of the final size relation across many different distributions of infectiousness~\cite{ma2006generality,miller:final}.

We are now in a position to explain why the inference in Figs.~\ref{fig:infer} and~\ref{fig:infer_more} resulted in narrow strips of plausible parameters.  The fact that the final size of epidemics depends only on the average of the offspring distribution makes it difficult to infer much about the disease other than $\Ro$ from the final size of larger outbreaks.  We can get a bit more information by observing how frequently outbreaks die out while still small, but it takes many outbreaks to collect enough data to observe this.  Additionally, because both the final size and the epidemic probability follow similar trends in the figures, it is difficult to isolate the disease parameters.  

Although it is difficult to identify the disease parameters from the final sizes of outbreaks, it should be noted that the reason for that is that the distribution of final outcomes are similar for these parameters.  This suggests two further questions:
\begin{itemize}
    \item Can we distinguish disease parameters from using the intermediate dynamics (that is, the number of infections at each ``generation'') instead of just the final size?
    \item Can we get similarly accurate predictions by using a family of offspring distributions that has only a single parameter?
\end{itemize}

\subsection{Parameter Inference from Intermediate Dynamics}
Now we assume that we observe the number infected in each generation.  Given $s_g$ susceptible and $i_g$ infected individuals, Eqn.~\eqref{eqn:Pigp1} shows that the probability of $m$ infections at the next generation is
\begin{align*}
P&(i_{g+1}=m| i_g, s_g, N)\\ &= \sum_{\ell=m}^\infty p_{s_g}(m, \ell) \left(\frac{s_g}{N}\right)^\ell \frac{1}{\ell!} \left.\diffm{\ell}{}{x} \left[ \frac{1-p}{1-px}\right]^{r i_g} \right|_{x = 1 - \frac{s_g}{N}}\\
&= 
[1-p]^{r i_g}\sum_{\ell=m}^\infty p_{s_g}(m, \ell) \left(\frac{s_g}{N}\right)^\ell \frac{1}{\ell!} \left.\diffm{\ell}{}{x} \left[ 1-px\right]^{-r i_g} \right|_{x = 1 - \frac{s_g}{N}}\\
&=[1-p]^{r i_g}\sum_{\ell=m}^\infty p_{s_g}(m, \ell) \left(\frac{s_g}{N}\right)^\ell \frac{(-p)^\ell(-ri_g)(-ri_g-1)\cdots(-ri_g-\ell+1)}{\ell!} \left[ 1-p\left(1 - \frac{s_g}{N}\right)\right]^{-r i_g-\ell}\\
&=\left[\frac{1-p}{1-p\left(1 - \frac{s_g}{N}\right)}\right]^{r i_g}\sum_{\ell=m}^\infty p_{s_g}(m, \ell) \left(\frac{ps_g}{N\left[ 1-p\left(1 - \frac{s_g}{N}\right)\right]}\right)^\ell \frac{(ri_g)(ri_g+1)\cdots(ri_g+\ell-1)}{\ell!}
\end{align*}

\begin{figure}
    \centering
    \includegraphics[width=\textwidth]{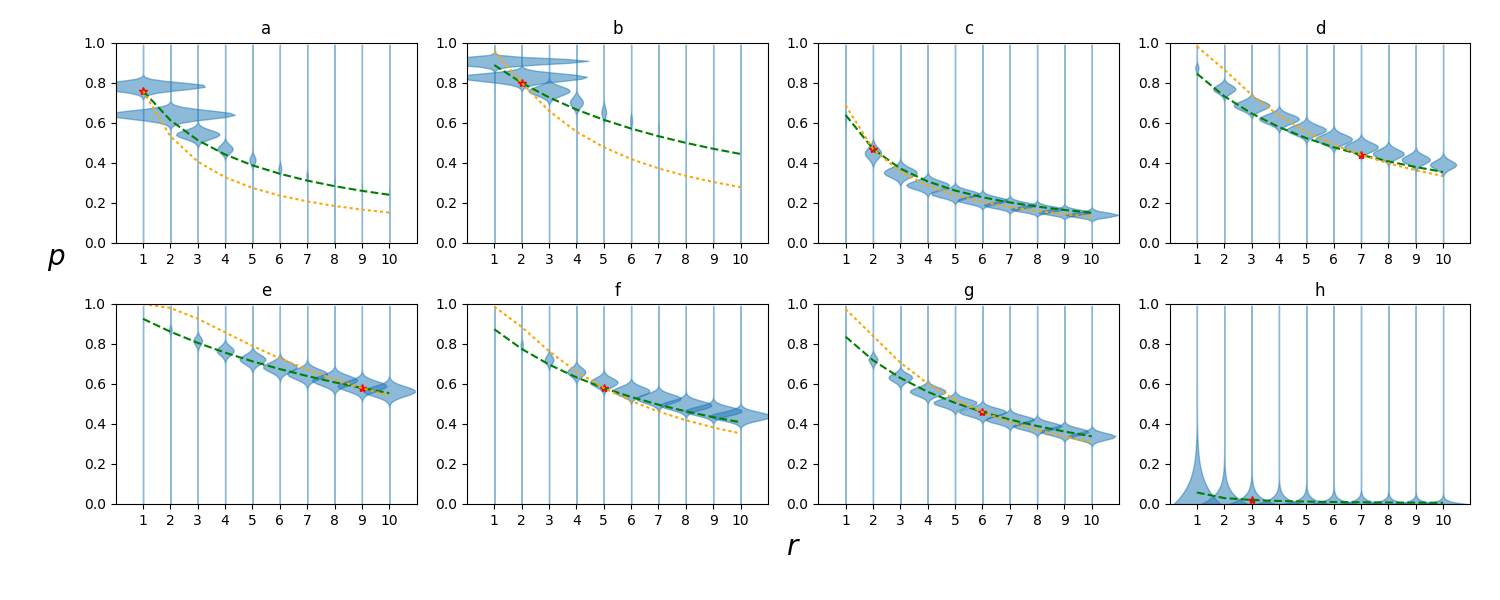}
      \caption{Scaled posterior probabilities for
      the same simulations used in Figure~\ref{fig:infer} (the outbreaks of Table~\ref{table:outbreaks}), but using the sizes of successive generations rather than just the final size.  The inference is able to improve the estimate, particularly for (e).  However, it still has difficulty distinguishing parameters with similar $\Ro$ or similar epidemic probabilities.}
    \label{fig:infer_dynamics}
\end{figure}

Figure~\ref{fig:infer_dynamics} shows the inferred parameter values based on the same simulations as Fig.~\ref{fig:infer}, but using the sizes of each generation.  The predictions are better for the same number of simulations, but they still struggle to distinguish between parameter values with similar means or similar epidemic probabilities.

\subsection{Assuming a Poisson offspring distribution}
We now investigate how our results change if we (incorrectly) assume that the data had been generated using a Poisson offspring distribution.  So our assumed family of offspring distributions (Poisson) does not match that of the actual family (negative binomial). Our goal is to see whether we can accurately infer $\Ro$ (which would correspond to predicting the final size of an epidemic in a large population).


If we assume a Poisson offspring distribution, then there is only a single parameter $\Ro$, the expected number of transmissions a single individual will cause.  Then $\mu(x) = e^{\Ro (x-1)}$ and our expression for the probability of $m$ infections at generation $g+1$ becomes
\begin{align*}
    P(i_{g+1}=m| i_g, s_g, N)
     &= \sum_{\ell=m}^\infty p_{s_g}(m,\ell) \left(\frac{s_g}{N}\right)^\ell \frac{1}{\ell!} \left.\diffm{\ell}{}{x} \left[e^{\Ro i_g(x-1)} \right]\right|_{x=1- \frac{s_g}{N}}\\
     &= \sum_{\ell=m}^\infty p_{s_g}(m,\ell) \left(\frac{s_g}{N}\right)^\ell \frac{1}{\ell!}(\Ro i_g)^{\ell} e^{\Ro i_g(- s_g/N)}
\end{align*}

\begin{figure}
    \centering
    \includegraphics[width=0.5\textwidth]{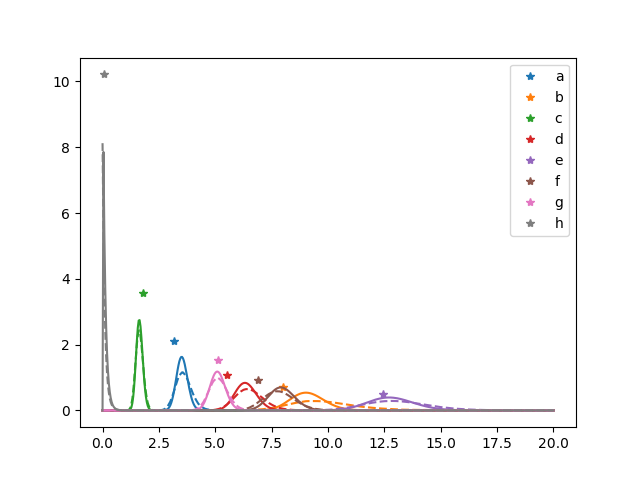}
    \caption{
    The inferred values of $\Ro$ as calculated assuming a Poisson distribution (solid) or Negative Binomial distribution (dashed) for the simulations of figure~\ref{fig:infer}.  The '*' denotes the actual location of $\Ro$.  The difference in prediction of the two distributions is not very large.}
    \label{fig:Poisson_R0}
\end{figure}
Figure~\ref{fig:Poisson_R0} shows a comparison of the probability density functions for $\Ro$ calculated assuming a Poisson distribution with the predictions for $\Ro$ found from Fig.~\ref{fig:infer}.  The predictions are generally consistent with one another, and relatively close to the true values.  

This suggests that we may get reasonable predictions using a simple Poisson distribution.  This contrasts somewhat with the observations of~\cite{lloyd2005superspreading} which focused on the impact of overdispersion.  This is in part because in our smaller populations it is difficult to observe superspreading events, but also because the main impact of overdispersion is on the probability of an epidemic.  With a relatively small number of observations it is difficult to measure the epidemic probability with high precision, and so many observations may be needed before we are able to observe overdispersion.  However, it seems reasonable that with a few observations in modest-sized populations, we can estimate the final size of an epidemic in the large-population limit.  When we assume homogeneous susceptibility in a well-mixed population, then only the mean of the offspring distribution ($\Ro$) affects the final size~\cite{ma2006generality,miller:final,miller:bounds}, which explains why we can accurately infer $\Ro$ even assuming the wrong distribution shape.  

\section{Discussion}

We have shown that given a known distribution of the number of transmissions an arbitrary individual will cause in a finite population of $N$ individuals, it is possible to calculate the size distribution of outbreaks.  The calculation is relatively efficient as it reduces to solving $\mat{C} \vec{q}=\vec{1}$ where $\mat{C}$ is a lower-triangular $N\times N$ matrix.  

Armed with this result we are able to perform inference to predict parameters of the offspring distribution in simulated epidemics.  Our predictions are reasonable, but there are some identifiability challenges.  In many of our tests, there is a large collection of plausible parameter sets, and these often correspond to parameters giving similar predictions for the reproductive number $\Ro$.  This is related to the fact that the final size of epidemics in the large population limit is a function only of $\Ro$, and not the specific offspring distribution.  This leads to the observation that it is relatively straightforward to predict $\Ro$ based on observations, but more improved predictions may require significantly more observations.  On the flip side, this implies that knowing $\Ro$ is often sufficient to give reasonable predictions about the range of outcomes.

\subsection{Weaknesses}
Unfortunately there are limitations to the applicability of our approach.  In particular, when the population has heterogeneous susceptibility, our approach here may give inaccurate conclusions.  It has been noted that in the large-population limit with homogeneous susceptibility only the average infectiousness influences the final size, and the heterogeneity only influences the probability of an epidemic.  In contrast, heterogeneity in susceptibility primarily impacts the final size and not the probability~\cite{miller:bounds,miller:final,miller:heterogeneity,ma2006generality}.  Thus if we are using the final size of large outbreaks to infer parameters of a model that incorrectly assumes homogeneous susceptibility, we may be taking an effect caused by heterogeneity in susceptibility and using it to infer information about the average infectiousness. 

\section*{Acknowledgment}
The visualization of the parameter inference in Figs~\ref{fig:infer}, \ref{fig:infer_more}, and \ref{fig:infer_dynamics} was created with help from Stack Overflow~\cite{stackoverflow:55886528}.  Michael Famulare provided significant valuable feedback.

\appendix
\section*{Appendix}
\section{A useful simple lemma}
Our derivations make  use of the following observation:

\begin{lemma}
Let a directed network $G$ whose nodes are labelled $1, \ldots, N$ be
given.  The nodes $1, \ldots, M$ have no edges to nodes in $M+1, \ldots, N$ if
and only if the out-component of node $1$ is entirely contained within
$1, \ldots, M$ and every node in $1, \ldots, M$ which is not in the
out-component of node $1$ has no edges to any node in $M+1, \ldots,
N$.
\end{lemma}

For notational simplicity, let $X$ denote the out-component of node
$1$ in the graph $G$.

First consider any directed graph for which $X$ contains at least
one node in $M+1, \ldots, N$.  Then we can choose a path from $1$ to a node in $M+1, \ldots, N$.
We take the first edge in that path that
reaches a node in $M+1, \ldots, N$.  It starts from a node in $1,
\ldots, M$.  So if $X$ contains a node in $M+1, \ldots, N$, then $G$
has at least one edge from a node in $1, \ldots, M$ to a node in $M+1,
\ldots, N$.  Consequently if there are no edges from any node in $1, \ldots, M$
to any node in $M+1, \ldots, N$, then $X$ must lie entirely in $1,
\ldots, M$.  Additionally if there are no edges from $1, \ldots, M$
to $M+1, \ldots, N$ then the nodes outside $X$ but within $1, \ldots,
M$ also have no edges to $M+1, \ldots, N$.

To show the other direction, we consider a directed graph for which $X$ lies entirely within
$1, \ldots, M$.  This means that there are no edges from nodes in $X$ to nodes
in $M+1, \ldots, N$ as otherwise those nodes would also lie in $X$.   If
the other nodes in $1, \ldots, M$ also have no
edges to nodes in $M+1, \ldots, N$ then there are no edges from $1,
\ldots, M$ to $M+1, \ldots, N$.

\bibliographystyle{plain}
\bibliography{finite_outbreak}

\end{document}